\documentclass[acmsmall,screen, nonacm]{acmart} 

\usepackage{amsmath}
\usepackage{algorithmicx}
\usepackage{algpseudocode}
\usepackage{algorithm}
\usepackage{graphicx}
\usepackage{textcomp}
\usepackage{xcolor}
\usepackage{listings}
\usepackage{booktabs}
\usepackage{multirow}
\usepackage{microtype}
\usepackage{array}
\usepackage{placeins}

\definecolor{sbtink}{HTML}{243447}
\definecolor{sbtblue}{HTML}{3478D4}
\definecolor{sbtteal}{HTML}{00A896}
\definecolor{sbtcoral}{HTML}{E85D5D}
\definecolor{sbtamber}{HTML}{F2B134}
\definecolor{sbtviolet}{HTML}{7A5AF8}

\usepackage{tikz}
\usetikzlibrary{shapes.geometric,arrows.meta,positioning,calc,fit}
\usepackage{pgfplots}
\pgfplotsset{compat=1.18}

\lstdefinelanguage{JavaScript}{
  keywords={break, case, catch, continue, debugger, default, delete, do, else, finally, for, function, if, in, instanceof, new, return, switch, this, throw, try, typeof, var, void, while, with, let, const, async, await},
  morecomment=[l]{//},
  morecomment=[s]{/*}{*/},
  morestring=[b]',
  morestring=[b]",
  sensitive=true
}

\setcopyright{none}

\begin{document}

\title[Story-Based Semantic Fuzzing]{From Stories to Bugs: Story-Based Semantic Fuzzing of Interactive Systems}


\author{Yeshayahu Weiss}
\affiliation{%
  \department{Faculty of Computer and Information Science}
  \institution{Ben-Gurion University of the Negev}
  \city{Be'er Sheva}
  \country{Israel}
}
\email{weissye@post.bgu.ac.il}

\author{Achiya Elyasaf}
\affiliation{%
  \department{Faculty of Computer and Information Science}
  \institution{Ben-Gurion University of the Negev}
  \city{Be'er Sheva}
  \country{Israel}
}
\email{achiya@bgu.ac.il}

\author{Gera Weiss}
\affiliation{%
  \department{Faculty of Computer and Information Science}
  \institution{Ben-Gurion University of the Negev}
  \city{Be'er Sheva}
  \country{Israel}
}
\email{geraw@bgu.ac.il}


\begin{abstract}
Complex interactive systems hide business-logic failures along two complementary axes: deep state-dependent behaviors that require long valid prefixes and concurrency-sensitive failures that depend on the ordering of otherwise legal operations. We present story-based semantic fuzzing, a methodology that treats semantically constrained compositions of behavioral stories as primary search objects. Implemented through Behavioral Programming, semantic binding and state constraints preserve legal prefixes, while event eligibility, schedule selection, and replay expose failure-relevant interleavings.
We evaluate the methodology on Magento and four stateful REST APIs. On Magento, overlap-aware ranking increased reproduction of a sporadic Shipping Fee race from $3.6\%$ to $41.2\%$; a controlled ordering intervention reduced a second failure signature from $100/100$ initial executions to $0/200$ constrained executions. For the REST study, a versioned end-to-end pipeline took each frozen public OpenAPI contract as its sole benchmark-specific model-generation input and automatically generated the Basic and Complex stories, dependency structure, and interface bindings. Under a shared black-box task and sequence-local evaluation contract, Basic and Complex confirmed $0/11$ and $11/11$ semantic fault classes, while RESTler and EvoMaster confirmed $2/11$ and $0/11$. A paired binding ablation produced $250/250$ valid dependent operations with binding and $0/250$ without it, while supplementary prefix controls localized the difference to preservation of legal relational state.
Together, the studies show that searching admissible semantic compositions provides a unified and effective representation for reaching deep states and exposing concurrency-sensitive business-logic failures.
\end{abstract}

\ccsdesc[500]{Software and its engineering~Software testing and debugging}
\ccsdesc[300]{Software and its engineering~Concurrency control}
\ccsdesc[300]{Software and its engineering~System testing}


\keywords{semantic fuzzing, story-based testing, stateful API testing, business-logic race conditions, concurrency testing, bug amplification, model-based testing, behavioral programming}

\maketitle

\section{Introduction}\label{sec:introduction}

Modern software systems are increasingly concurrent and stateful. In domains ranging from e-commerce to enterprise Software as a Service (SaaS), business behavior emerges from interactions among multiple user sessions, background services, roles, and long sequences of dependent operations~\cite{lu2008learning}. Although individual workflows may appear simple, their combinations create a large space of system states and execution orders~\cite{clarke2012model}. Critical defects can therefore remain hidden along two complementary dimensions: deep state-dependent behaviors that require long valid prefixes, and concurrency-sensitive failures whose manifestation depends on the relative timing of otherwise valid operations~\cite{fonseca2010study}.

Prior work establishes each difficulty separately. Systematic concurrency testing controls schedules to expose and reproduce order-dependent failures~\cite{lu2008learning,musuvathi2008chess}, while stateful REST API fuzzing constructs dependent request sequences to reach deeper service states and documents the rapid growth of the sequence space~\cite{atlidakis2019restler,lyu2023miner}. We connect these difficulties through semantic reachability: a failure remains hidden when testing loses either the valid state-building prefix or the failure-inducing ordering of otherwise valid operations.

In this paper, we use the term \emph{business-logic race condition} to denote an externally observable application-level semantic violation whose manifestation depends on the relative timing or interleaving of valid operations across concurrent actors or sessions. This notion is intentionally broader than a code-level \emph{data race}. A business-logic race condition can be observed as incorrect application behavior without determining whether the internal cause is a shared-memory race, transaction ordering, stale state, synchronization, caching, or another implementation mechanism. This black-box perspective is important for distributed and web systems, where the tester often observes only externally visible operations and outcomes.

Existing automated testing techniques address important parts of this problem. Stateful API fuzzers construct multi-request tests by exploiting interface dependencies, search objectives, or execution feedback, substantially extending testing beyond isolated requests~\cite{atlidakis2019restler,arcuri2025tool,lin2023forest,rooijakkers2025wuppiefuzz}. Other black-box approaches enrich schema-driven generation with explicit dependencies, constraints, or generic behavioral properties~\cite{martinlopez2020restest,karlsson2024behaviours}. However, preserving a difficult business-level prefix may require domain-specific information that is not explicit in the interface schema alone, such as which entity created earlier must be reused later, which relationship between entities must remain valid, which state transition is admissible, or which business invariant determines failure. Similarly, exposing a business-logic race condition may require exploring the relative scheduling of meaningful operations across multiple actors rather than generating another valid request sequence.

We address these challenges through \emph{story-based semantic fuzzing}, a methodology for transforming business-level behavioral stories into an executable search space of semantically admissible compositions~\cite{harel2012behavioral,harel2011model}. The methodology separates behavioral intent from system interaction, preserves entities and relations through shared semantic context and binding, constrains compositions to legal executions, and evaluates outcomes with business-level oracles. It explores this space in two complementary ways: composing stories into long, state-building prefixes and varying the relative order and overlap of concurrently active stories.

Provengo~\cite{barsinai2023provengo} operationalizes the methodology by providing the Behavioral Programming runtime for composing executable stories and constraints, while API and Selenium bindings connect high-level events to the SUT. The methodology supplies the scenario-construction rules, semantic context and binding, constraints, and class-specific oracles that determine which compositions preserve business meaning. Within the resulting model, semantic binding preserves state-building prefixes, while the BP scheduler generates admissible interleavings by repeatedly selecting requested events that are not blocked. Provengo's sampling and ranking facilities support practical selection from this space for execution against the SUT, while targeted replay amplifies observed failure-prone schedules. The resulting search object--an admissible story composition--therefore supports both dimensions evaluated in this paper.

These capabilities correspond to two axes of the verification problem. The \emph{interleaving axis} concerns the relative ordering and overlap of operations across actors or sessions. In Magento, for example, overlapping cart updates and checkout activity across sessions can expose business-logic race conditions that occur only under particular schedules. Once such a failure is observed, overlap-aware ranking and replay can prioritize schedules that exhibit the failure characteristics and increase the probability of its reproduction. The \emph{depth axis} concerns long semantically valid prefixes. In stateful REST APIs, reaching a target operation may require creating the correct entities, preserving relationships among them, and carrying the resulting identifiers through several dependent requests before a business invariant can even be exercised. Story-based semantic guidance makes these relationships explicit and allows generated executions to remain semantically admissible as they explore deeper states. Thus, the two axes involve different manifestations of the same underlying challenge: preserving the semantic conditions that make a difficult failure reachable.

To evaluate how story-based semantic fuzzing addresses these two dimensions, we organize the study around three research questions:
\begin{itemize}
\item \textbf{RQ1 (The Interleaving Axis):} Can story-based semantic fuzzing characterize schedule sensitivity in multi-session business-logic failures through controlled ordering, and, for sporadic manifestations, can overlap-aware ranking amplify reproduction relative to unguided replay?
\item \textbf{RQ2 (The Depth Axis):} To what extent does automatic, parameterized scenario construction preserve the long, valid prefixes required to expose deep business-logic faults? We evaluate its effect through Basic and Complex profiles generated from the same frozen OpenAPI by the same pipeline, and assess external relevance against schema-driven black-box API fuzzers under a shared sequence-local evaluation contract.
\item \textbf{RQ3 (Cross-Axis Mechanism):} How does a common representation based on semantically admissible story compositions support both long-prefix reachability and interleaving-sensitive failure reproduction, and which mechanisms account for the observed effects?
\end{itemize}

This paper makes three contributions. First, we introduce story-based semantic fuzzing, a methodology that transforms behavioral requirements into executable spaces of admissible story compositions for exploring long state-building prefixes and cross-session interleavings. Second, we evaluate both dimensions. On Magento, selected schedules increased reproduction of the sporadic Shipping Fee race from $3.6\%$ to $41.2\%$, while a separate Empty Cart intervention assessed ordering sensitivity. Across four REST APIs, automatically generated Basic and Complex profiles of the same pipeline confirmed 0/11 and 11/11 semantic fault classes from the same frozen public inputs, isolating the contribution of parameterized composition richness. Common-task comparisons with RESTler and EvoMaster assess external schema-driven methods under shared contracts, SUTs, and sequence-local confirmation rules. Third, we provide a reproducible artifact containing both SBT profiles, frozen inputs, generator provenance, retained executions, generated tests, sequence-local evaluators, canonical evidence, and replay and analysis scripts.

\section{Related Work}\label{sec:related}
Our work intersects with several established domains in software testing: model-based testing (MBT), black-box API fuzzing, concurrency heuristics, bug amplification, and test oracles.

\subsection{Model-Based Testing (MBT)}
MBT derives executable tests from FSMs, statecharts, and other behavioral models~\cite{bochmann1977unified,harel1987statecharts}. Industrial studies report coverage benefits alongside modeling and CI/CD adoption costs~\cite{garousi2021model,cao2025mixed,lonetti2023model}. Behavioral Programming instead composes partial stories without requiring a complete global state machine~\cite{harel2012behavioral}, making multi-user behavior incrementally modelable while retaining explicit semantic guidance.

\subsection{Black-box API Testing and Behavioral API Generation}
Stateful API fuzzers build multi-request tests from inter-request dependencies: RESTler infers producer-consumer relationships~\cite{atlidakis2019restler}, EvoMaster evolves sequences toward coverage objectives~\cite{arcuri2025tool}, foREST prioritizes API dependency trees~\cite{lin2023forest}, Wuppie Fuzz mutates sequences using coverage feedback~\cite{rooijakkers2025wuppiefuzz}, and Schemathesis derives schema-based properties~\cite{hatfield2022deriving}. Other black-box approaches enrich schema-driven generation with additional constraints or behavioral knowledge. RESTest supports explicit inter-parameter dependencies and constraint-based generation of valid requests~\cite{martinlopez2020restest}, while Karlsson et al.\ generate multi-call examples from generic REST behavioral properties to move testing closer to application behavior and business logic~\cite{karlsson2024behaviours}. These approaches demonstrate the value of semantics beyond raw schema structure. Our method uses analyst-authored, domain-specific business invariants and explicit cross-request entity binding to constrain the search space, while treating modular story compositions and cross-session interleavings as first-class search objects.

\subsection{Concurrency Heuristics and Noise Injection}
Schedule-manipulation heuristics expose races in internal program threads~\cite{eytani2003heuristics,krena2012noise}. Our black-box method moves scheduling to the story level, where distinct users interact concurrently and expose architectural business-logic race conditions without internal instrumentation.

\subsection{Search-Based Amplification}
Search-based testing uses coverage or fault-detection objectives to steer exploration~\cite{formica2023search}, while Test Amplification adapts tests to exercise new behavior~\cite{danglot2019automatic}. Black-box bug amplification ranks and replays executions around an observed sporadic failure~\cite{weiss2025black}. We embed this concurrency-specific mechanism in a broader semantic fuzzing process whose search objects are constrained story compositions and whose prefixes also reach sequential stateful-API failures.

\subsection{Oracles and Static Analysis}\label{sec:oracles-static}
Logic bugs require state-based oracles beyond generic HTTP status checks~\cite{taromirad2025assertions}; we integrate such assertions into the test-side context, consistent with guidance on automatable oracles~\cite{meszaros2007xunit}. Static race analyzers such as RacerD and Goblint reason about code-level data races and synchronization properties~\cite{gorogiannis2019racerd,vojdani2016goblint}; by design, such source-level analyses do not directly encode the distributed, application-level business invariants targeted by our black-box story models.

In this paper, a \emph{semantic fault class} denotes a distinct business invariant whose violation requires a legal multi-request prefix. A \emph{semantic failure} is an observed execution of such a violation. We count a semantic fault class as confirmed only when the executed prefix satisfies its declared preconditions and a class-specific oracle confirms the invariant violation; malformed messages, unexpected status codes, schema mismatches, timeouts, and generic server errors alone do not qualify. We call the valid setup a \emph{semantic prefix}, and the reuse of concrete entities and relations created along that prefix \emph{semantic binding}.

\subsection{Positioning of This Paper}
Behavioral Programming architectures provide the execution foundation~\cite{weiss2021testing}, Provengo supplies a practical story-composition runtime~\cite{barsinai2023provengo}, and black-box bug amplification supplies ranking and targeted replay for sporadic business-logic race conditions~\cite{weiss2025black}. Our contribution lies in the search formulation built over these foundations: story-based semantic fuzzing treats admissible, semantically constrained story compositions as first-class search objects, preserves cross-request entity relationships through semantic binding, and uses the same representation to explore both long stateful prefixes and cross-session interleavings. Overlap-aware amplification is one mechanism within this broader formulation. ~\autoref{tab:paradigms} summarizes this positioning.

\begin{table*}[t]
\centering
\caption{Representative comparison of testing paradigms relevant to this work. Static analysis reasons over code-level properties, while stateful API fuzzing constructs request sequences to improve structural and state reachability. Story-Based Testing (SBT) instead makes semantically constrained story compositions the primary search object, enabling explicit reasoning about deep semantic prefixes and cross-session interleavings.}
\label{tab:paradigms}
\small
\begin{tabular}{@{}
>{\raggedright\arraybackslash}p{0.155\textwidth}
>{\raggedright\arraybackslash}p{0.225\textwidth}
>{\raggedright\arraybackslash}p{0.265\textwidth}
>{\raggedright\arraybackslash}p{0.265\textwidth}
@{}}
\toprule
\textbf{Feature} & \textbf{Static Analysis} & \textbf{Stateful Fuzzing} & \textbf{Story-Based Testing} \\
& (\textit{e.g., RacerD, Goblint}) & (\textit{e.g., RESTler, EvoMaster}) & (\textit{Ours}) \\
\midrule
\textbf{Representative Goal} & Memory Safety \& Data Races & Structural/State Reachability \& Fault Exposure & Semantic Logic \& Scheduling \\
\textbf{Search Object} & Code-Level Program Representation & Stateful Request Sequences / Dependency Chains & Admissible Story Compositions \\
\textbf{Depth Axis} & Static Program Reasoning & Dependency-Guided Sequencing & Explicit Semantic Prefixes \\
\textbf{Interleaving Axis} & Program-Level Concurrency Analysis & Sequential API Request Sequences & Cross-Session Interleavings \& Amplification \\
\textbf{Input Model} & Source Code & OpenAPI / Schema & Stories, Constraints \& Semantic Context \\
\textbf{Oracle Scope} & Static Properties & Interface/Contract \& Generic Failure Signals & Explicit State-Based Business Invariants \\
\bottomrule
\end{tabular}
\end{table*}

\section{Background}\label{sec:background}

\subsection{Story-Based Testing (SBT)}
Story-Based Testing, rooted in Scenario Testing~\cite{kaner2003introduction}, represents user behavior as modular executable narratives. Hereafter, SBT denotes Story-Based Testing; story-based semantic fuzzing denotes the broader methodology introduced here. A story is a behavioral thread (b-thread) that captures business intent through a coherent sequence of high-level events, distinct from an Agile user story~\cite{getto2020story,masud2017automated}. Independent stories compose at run time rather than forming fixed linear tests; for example, shopper and administrator stories can interleave to expose interactions that neither story contains alone. Anti-stories declaratively block illegal combinations, such as issuing an administrator-only action from a session lacking that role, thereby pruning invalid executions without modifying the positive stories~\cite{barsinai2019scenario}. Coverage criteria and optimization then select compact, high-value suites from the remaining admissible space.

\subsection{Overview of Behavioral Programming (BP)}
At the core of the proposed approach lies the BP paradigm~\cite{harel2012behavioral,harel2010programming}, implemented via the BPjs~\cite{barsinai2018bpjs} library. In BP, an executable model consists of concurrent, independent behavioral fragments called b-threads (or stories in our context)~\cite{elyasaf2021context}. At each synchronization point, each story declares three event sets: requested events ($R$), waited-for events ($W$), and blocked events ($B$). An application-agnostic execution engine selects an event $e \in (\bigcup R) \setminus (\bigcup B)$, resumes the threads waiting for or requesting $e$, and blocks forbidden execution paths, repeating this cycle until no further events can occur; see ~\autoref{fig:bp_cycle}.

Consider two independent stories requesting three presses of a green button and 10 presses of a red button. Without constraints, the engine explores all admissible interleavings, yielding $\binom{13}{3} = 286$ distinct runs. If a constraint story blocks a green press immediately after another green press, the system coordinates the concurrent threads, and the number of admissible runs drops to $\binom{11}{3} = 165$ without internal code modifications or changes to the original stories.

\subsection{The Provengo Tool}
Provengo~\cite{barsinai2023provengo} is the BP-based Scenario-Driven Model-Based Testing platform on which we implement the method. It provides the runtime for composing executable stories, constraint stories, and operational bindings~\cite{harel2012behavioral,barsinai2018bpjs}; the semantic fuzzing strategy introduced here is layered on that infrastructure.

\textbf{Modeling:} Stories capture high-level events (e.g., \texttt{Login}, \texttt{AddToCart}, \texttt{Checkout}), while bindings expand those events into low-level Selenium steps or API calls. This separation keeps intent stable as execution technology changes, akin to behavior-driven abstractions~\cite{wynne2017cucumber}. Sessions are logical execution containers: a story may span sessions, and several stories may influence one session, enabling multi-role concurrent workflows.

\textbf{Generation and execution:} Given a story-and-constraint model, Provengo enumerates admissible interleavings via BP's event-selection loop. In addition to the BP runtime, the tool provides \texttt{sample} and \texttt{ensemble} commands for selecting diverse candidate suites. The selected suites are then executed against the SUT (\texttt{run}), and the outcomes are summarized (\texttt{report}). These steps correspond to the mid-to-late pipeline phases that turn a stable model into concrete suites and results.

\textbf{High-level events:} Provengo interleaves high-level events across sessions and then expands them into bound action sequences. Constraint stories prune illegal or irrelevant alternatives, while the remaining event orderings represent admissible multi-session interference. This preserves behavioral intent without entangling it with UI or API mechanics; ~\autoref{fig:events} illustrates the mapping.

\begin{figure}[t]
\centering
\begin{minipage}[t]{0.48\textwidth}
\vspace{0pt}
\centering
\begin{minipage}[t][4cm][t]{\linewidth}
\centering
\resizebox{\linewidth}{!}{%
\begin{tikzpicture}[
    font=\sffamily\scriptsize,
    >={Stealth[length=4pt,width=5pt]},
    link/.style={->, draw=sbtink!82, line width=0.6pt},
    card/.style={rounded corners=2pt, draw=sbtink!20, fill=white,
                 minimum width=1.55cm, minimum height=0.46cm},
    chip/.style={rounded corners=4pt, text=white,
                 font=\sffamily\tiny\bfseries,
                 minimum width=0.88cm, minimum height=0.35cm, inner sep=1pt}
]
    \node[card, xshift=2pt, yshift=2pt] (storyback) at (0,1.25) {};
    \node[card] (stories) at (0,1.25) {ACTIVE STORIES};

    \node[chip, fill=sbtblue!48, xshift=4pt, yshift=4pt] (req3) at (2.05,1.75) {};
    \node[chip, fill=sbtblue!72, xshift=2pt, yshift=2pt] (req2) at (2.05,1.75) {};
    \node[chip, fill=sbtblue] (req) at (2.05,1.75) {REQUEST};
    \node[chip, fill=sbtcoral!58, xshift=3pt, yshift=3pt] (block2) at (2.05,0.75) {};
    \node[chip, fill=sbtcoral] (block) at (2.05,0.75) {BLOCK};
    \node[chip, fill=sbtteal, minimum width=1.05cm] (wait) at (0,0.42) {WAIT FOR};

    \node[circle, fill=sbtink, text=white, minimum size=1.27cm,
          font=\sffamily\scriptsize\bfseries, align=center] (engine) at (3.75,1.25)
          {BP\\ENGINE};
    \node[circle, fill=sbtamber, text=sbtink, minimum size=0.82cm,
          font=\sffamily\small\bfseries] (event) at (5.70,1.25) {$e_k$};

    \draw[link] (stories.east) -- (req.west);
    \draw[link] (stories.east) -- (block.west);
    \draw[-, draw=sbtteal, line width=0.8pt] (stories.south) -- (wait.north);
    \draw[link] (req.east) -- (engine.north west);
    \draw[link] (block.east) -- (engine.south west);
    \draw[link, line width=0.85pt] (engine.east) --
        node[above=3pt, text=sbtink, align=center,
             font=\sffamily\tiny\bfseries]
        {requested and\\not blocked} (event.west);
    \draw[->, draw=sbtblue, line width=0.95pt, rounded corners=4pt]
        (event.north) -- (5.70,2.24) --
        node[midway, above=2pt, text=sbtblue,
             font=\sffamily\tiny\bfseries]
        {notify requesters and waiters}
        (0,2.24) -- (stories.north);
\end{tikzpicture}%
}
\end{minipage}
\caption{Behavioral Programming (BP) event selection. At each synchronization point, active stories may request, wait for, or block events. The engine selects a requested event that is not blocked and then notifies the stories that requested or waited for the selected event.}
\label{fig:bp_cycle}
\end{minipage}\hfill
\begin{minipage}[t]{0.48\textwidth}
\vspace{0pt}
\centering
\begin{minipage}[t][4cm][t]{\linewidth}
\centering
\resizebox{\linewidth}{!}{%
\begin{tikzpicture}[
    font=\sffamily\scriptsize,
    >={Stealth[length=4pt,width=5pt]},
    storyA/.style={rounded corners=5pt, fill=sbtblue, text=white,
                   minimum width=0.94cm, minimum height=0.38cm,
                   font=\sffamily\tiny\bfseries},
    storyB/.style={rounded corners=5pt, fill=sbtcoral, text=white,
                   minimum width=0.94cm, minimum height=0.38cm,
                   font=\sffamily\tiny\bfseries},
    actionA/.style={circle, fill=sbtblue, text=white, minimum size=0.47cm,
                    font=\sffamily\tiny\bfseries, inner sep=0pt},
    actionB/.style={circle, fill=sbtcoral, text=white, minimum size=0.47cm,
                    font=\sffamily\tiny\bfseries, inner sep=0pt},
    bindingA/.style={->, draw=sbtblue!82!sbtink, line width=0.62pt},
    bindingB/.style={->, draw=sbtcoral!84!sbtink, line width=0.62pt}
]
    \node[text=sbtblue, font=\sffamily\tiny\bfseries] at (1.85,2.95) {SESSION A};
    \draw[sbtblue!45, line width=0.7pt] (0.45,2.72) -- (3.25,2.72);
    \node[storyA] (a1) at (0.82,2.35) {STORY 1};
    \node[storyA] (a2) at (1.85,2.35) {STORY 2};
    \node[storyA] (a3) at (2.88,2.35) {STORY 3};

    \node[text=sbtcoral, font=\sffamily\tiny\bfseries] at (4.88,2.95) {SESSION B};
    \draw[sbtcoral!50, line width=0.7pt] (3.70,2.72) -- (6.05,2.72);
    \node[storyB] (b1) at (4.25,2.35) {STORY 4};
    \node[storyB] (b2) at (5.45,2.35) {STORY 5};

    \node[anchor=east, text=sbtink, align=right] at (0.18,0.38) {Bound\\execution};
    \draw[->, draw=sbtink!78, line width=0.75pt] (0.35,0.38) -- (6.20,0.38);
    \node[actionA] (x1) at (0.65,0.38) {$a_{1,1}$};
    \node[actionB] (x2) at (1.30,0.38) {$b_{2,4}$};
    \node[actionA] (x3) at (1.95,0.38) {$a_{3,3}$};
    \node[actionB] (x4) at (2.60,0.38) {$b_{4,5}$};
    \node[actionA] (x5) at (3.25,0.38) {$a_{5,2}$};
    \node[actionB] (x6) at (3.90,0.38) {$b_{6,4}$};
    \node[actionA] (x7) at (4.55,0.38) {$a_{7,3}$};
    \node[actionB] (x8) at (5.20,0.38) {$b_{8,5}$};
    \node[actionB] (x9) at (5.85,0.38) {$b_{9,4}$};

    \draw[bindingA] (a1.south) to[out=-88,in=96] (x1.north);
    \draw[bindingA] (a2.south) to[out=-78,in=100] (x5.north);
    \draw[bindingA] (a3.south) to[out=-112,in=92] (x3.north);
    \draw[bindingA] (a3.south) to[out=-62,in=98] (x7.north);
    \draw[bindingB] (b1.south) to[out=-132,in=86,looseness=1.05] (x2.north);
    \draw[bindingB] (b1.south) to[out=-104,in=92] (x6.north);
    \draw[bindingB] (b1.south) to[out=-58,in=98] (x9.north);
    \draw[bindingB] (b2.south) to[out=-128,in=86] (x4.north);
    \draw[bindingB] (b2.south) to[out=-78,in=96] (x8.north);

    \draw[draw=sbtviolet, line width=1.25pt, rounded corners=2pt]
        (0.42,-0.08) -- (6.08,-0.08);
    \node[text=sbtviolet, font=\sffamily\tiny\bfseries] at (3.25,-0.42)
        {ONE ADMISSIBLE INTERLEAVING};
    \node[text=sbtink!82, font=\sffamily\tiny] at (3.25,-0.76)
        {$a/b$: session; first index: execution position; second index: source story};
\end{tikzpicture}%
}
\end{minipage}
\caption{Example binding of two sessions into one admissible execution. Stories expand into bound actions that may interleave across stories; repeated actions need not remain adjacent. Labels encode execution position and source story; the $a/b$ prefix identifies the session.}
\label{fig:events}
\end{minipage}
\Description{Left: diagram of the Behavioral Programming synchronization cycle, where active stories provide requested, blocked, and waited-for events to the BP engine, which selects a requested event that is not blocked and notifies the relevant stories. Right: diagram showing five stories distributed across two sessions and mapped to nine actions in one admissible interleaving; action labels identify the originating session, execution position, and source story.}
\end{figure}

\subsection{Empirical Baselines: RESTler and EvoMaster}
We compare SBT with RESTler~\cite{atlidakis2019restler} and EvoMaster~\cite{arcuri2025tool}, two representative black-box API-testing tools. All three methods receive the same public OpenAPI operations and exercise the same SUT boundary. SBT automatically derives its stories, dependencies, and bindings, while RESTler and EvoMaster apply their native OpenAPI transformations. The evaluation therefore compares end-to-end semantic reachability from the same public contract.

\section{Method: Story-Based Semantic Fuzzing}\label{sec:method}
Our method transforms business-level requirements into an executable search model of semantically admissible story compositions. In the general workflow and the Magento study, analysts define modular stories and anti-stories, separate high-level behavioral intent from operational bindings, preserve created entities and relations in a shared semantic context, and specify business-level oracles. In the REST study, the OpenAPI-to-SBT pipeline automatically derives the corresponding executable stories, dependencies, and bindings from the public contract; the semantic evaluators remain external to generation. Together, these elements determine which executions remain meaningful as exploration varies story composition.

Exploration then specializes to the failure dimension. For depth-sensitive faults, semantic binding and constraints preserve the long state-building prefix required before a target operation becomes meaningful. For concurrency-sensitive faults, multiple stories remain active concurrently and coordinate through requested, waited-for, and blocked events. At each step, the BP scheduler selects a requested event that is not blocked; alternative eligible selections produce different admissible schedules. Provengo's sampling and ranking facilities make it practical to select a compact, high-value subset of these schedules for execution against the SUT, while targeted replay is used subsequently to amplify observed failure-prone schedules.

Across both axes, the incremental spiral lifecycle (\autoref{fig:spiral}) iteratively: (1) authors and stabilizes the executable model; (2) samples admissible compositions; (3) selects and executes traces against the SUT; (4) applies semantic oracles; and (5) replays confirmed or high-risk candidates where appropriate. Provengo supplies the Behavioral Programming runtime and the operational infrastructure for this process. The resulting procedure searches over admissible story compositions, as formalized in~\autoref{alg:fuzzing_loop}.

\begin{algorithm}[t]
\caption{Story-based semantic fuzzing and amplification loop. The procedure samples admissible story compositions, applies semantic bindings, executes and records the resulting traces, retains oracle-confirmed failures, and amplifies traces that either violate an oracle or exceed the configured hazard-score threshold.}
\label{alg:fuzzing_loop}
\begin{algorithmic}[1]
\Require Stories $\mathcal{S}$, Anti-stories $\mathcal{A}$, Operational Bindings $\mathcal{B}$, Semantic Context $\mathcal{C}$, Oracles $\mathcal{O}$, Hazard Templates $\mathcal{H}$, Threshold $\tau$
\Ensure Executed traces $\mathcal{E}$, confirmed failures $\mathcal{F}$, amplified replay evidence $\mathcal{R}$
\State $\mathcal{E} \leftarrow \emptyset$, $\mathcal{F} \leftarrow \emptyset$, $\mathcal{R} \leftarrow \emptyset$
\While{testing budget remaining}
    \State $\mathcal{T} \leftarrow \text{SampleAdmissibleComposition}(\mathcal{S}, \mathcal{A}, \text{BP\_Scheduler})$
    \State $\mathcal{T}' \leftarrow \text{ApplySemanticBindings}(\mathcal{T}, \mathcal{B}, \mathcal{C})$
    \State $L \leftarrow \text{ExecuteAndRecord}(\mathcal{T}', \text{SUT}, \mathcal{O})$
    \State $\mathcal{E} \leftarrow \mathcal{E} \cup \{L\}$
    \State $h \leftarrow \text{EvaluateTemplates}(L, \mathcal{H})$
    \If{$\text{Violated}(\mathcal{O}, L)$}
        \State $\mathcal{F} \leftarrow \mathcal{F} \cup \{L\}$
    \EndIf
    \If{$h \ge \tau \lor \text{Violated}(\mathcal{O}, L)$}
        \State $\mathcal{R} \leftarrow \mathcal{R} \cup \text{AmplifyCandidate}(\mathcal{T}', \text{scheduling\_variation})$
    \EndIf
\EndWhile
\State \Return $\mathcal{E}, \mathcal{F}, \mathcal{R}$
\end{algorithmic}
\end{algorithm}

\subsection{Method Inputs}
The method receives executable stories $\mathcal{S}$, admissibility constraints $\mathcal{A}$, operational bindings $\mathcal{B}$, semantic context $\mathcal{C}$, semantic oracles $\mathcal{O}$, and hazard templates $\mathcal{H}$. Their modeling semantics, including high-level events, sessions, requested/waited-for/blocked sets, anti-stories, and abstract-to-concrete binding, are defined in~\autoref{sec:background}. The method uses these inputs to sample admissible compositions, preserve state across their prefixes, rank hazardous traces, execute them on the SUT, and amplify failure-prone schedules.

\subsection{Incremental Story-Based Testing Lifecycle}
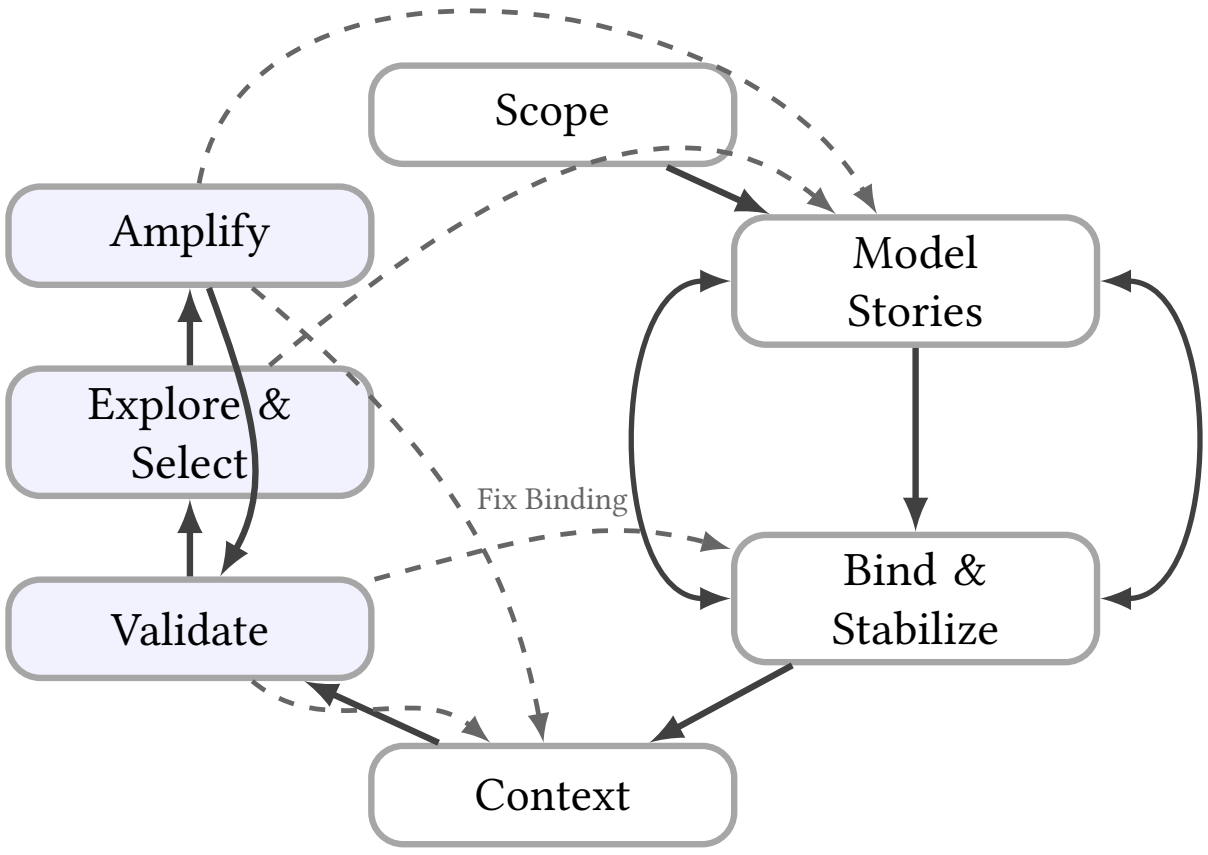
\begin{figure}[t]
\centering
\begin{tikzpicture}[
    font=\small,
    box/.style={draw=gray!70, rounded corners=6pt, line width=1.2pt, fill=blue!5, text centered, minimum width=2.4cm, minimum height=0.65cm},
    plainbox/.style={draw=gray!70, rounded corners=6pt, line width=1.2pt, fill=white, text centered, minimum width=2.4cm, minimum height=0.65cm},
    ar/.style={->, >=latex, line width=1.2pt, color=black!75},
    biar/.style={<->, >=latex, line width=1.0pt, color=black!75},
    dashar/.style={->, >=latex, dashed, line width=0.9pt, color=black!60}
]
    \node[plainbox] (scope) at (0, 3.0) {Scope};
    \node[plainbox, text width=2cm, align=center] (model) at (2.4, 1.9) {Model\\Stories};
    \node[plainbox, text width=2cm, align=center] (bind) at (2.4, -0.2) {Bind \&\\Stabilize};
    \node[plainbox] (context) at (0, -1.5) {Context};
    
    \node[box] (amplify) at (-2.4, 2.2) {Amplify};
    \node[box, text width=2cm, align=center] (explore) at (-2.4, 0.9) {Explore \&\\Select};
    \node[box] (validate) at (-2.4, -0.4) {Validate};
    
    \draw[ar] (scope) -- (model);
    \draw[ar] (model) -- (bind);
    \draw[ar] (bind) -- (context);
    \draw[ar] (context) -- (validate);
    \draw[ar] (validate) -- (explore);
    \draw[ar] (explore) -- (amplify);
    \draw[ar] (amplify) to[out=290, in=60] (validate);
    
    \draw[biar] (model.east) to[out=0, in=0] (bind.east);
    \draw[biar] (model.west) to[out=180, in=180] (bind.west);
    
    \draw[dashar] (validate) to[out=15, in=165] node[above, font=\tiny, text=gray!80!black] {Fix Binding} (bind);
    \draw[dashar] (explore) to[out=40, in=140] (model);
    \draw[dashar] (amplify) to[out=80, in=120] (model);
    \draw[dashar] (validate) to[out=320, in=140] (context);
    \draw[dashar] (amplify) to[out=320, in=100] (context);
\end{tikzpicture}
\caption{The incremental story-based testing lifecycle. Solid arrows show the main progression from scope and modeling through binding, context, validation, exploration and selection, and amplification. Feedback loops support fixing bindings, refining the model and context, and adding new stories as testing scope expands.}
\label{fig:spiral}
\Description{Flow diagram with Scope and Model Stories on the right, Context and Validate below, and Explore and Select and Amplify on the left. Solid arrows form the main execution cycle, while dashed arrows return validation and amplification evidence to bindings, context, and the model.}
\end{figure}

In the general workflow, analysts may supply stories, constraints, bindings, context, oracles, and hazard templates; the framework then samples, selects, executes, and replays admissible compositions (see~\autoref{fig:spiral}). For the REST study, the OpenAPI-to-SBT pipeline instead generates the executable model automatically from the frozen contract. In Library, Garage, and Pharmacy, Basic uses 1/1 entity/action multiplicities and Complex uses 5/7. In NetBox, both use 5/7, isolating direct generation from dependency-derived composition. All generated inputs, parameters, and artifacts are retained with hashes.

\textbf{Modeling phase:} We begin by defining the test vocabulary and composing an initial set of stories, typically by separating distinct user sessions to maximize interleaving. To support data-driven testing, stories are parameterized using a shared context. Instead of hardcoding specific product data, a story requests resources from the context, allowing the same logic to cover a diverse range of inventory items. Queries expose dynamic subsets of the context, enabling the selection of entities (e.g., active users or available products) for test execution.

A key benefit of this compositional lifecycle is that the complete behavioral model need not be defined before testing begins. Analysts can model and execute one functional slice, such as shopping and checkout, and subsequently add independent stories for administrative actions, additional stores, payment methods, or other behaviors without rewriting the existing stories. The executable model therefore evolves incrementally as testing scope expands, in contrast to MBT approaches that require substantial portions of a global behavioral model to be established before meaningful execution can begin.

\textbf{Stabilization phase:} Before large-scale interleaving exploration, the model undergoes stabilization. Short preliminary runs against the live SUT validate bindings, reveal order-sensitive interactions, and filter out actuator noise, such as broken CSS selectors or API authentication errors. If a binding fails, the workflow iterates between binding definition and validation until the component executes reliably.

\textbf{Exploration phase:} Once the model is stable, we proceed to exploration. Directly executing all possible interleavings on the SUT is computationally prohibitive because of I/O latency. Instead, we perform offline sampling, in which the BP execution engine generates thousands of admissible traces in a high-speed abstract simulation. This process produces a large pool of valid test scenarios spanning varied event orderings and session overlaps. To surface realistic overlaps, we separate add-to-cart flows from checkout flows and run them in distinct sessions. We create multiple add-to-cart stories (each selecting a product) and one checkout story, prompting the engine to explore plausible multi-session interleavings.

Unconstrained compositions can include orderings that violate domain preconditions as well as orderings that remain behaviorally meaningful but expose schedule-sensitive outcomes. We distinguish two uses of constraints: pruning genuinely inadmissible traces and, experimentally, holding a selected ordering relation fixed to assess its effect on an observed manifestation. If conflicting constraints lead to a model-level deadlock, the Provengo engine detects and reports it during offline sampling, allowing the tester to resolve it before live execution.

\textbf{Ensemble selection:} From this generated pool, we use an overlap-aware ranking algorithm to select a compact ensemble of tests. The ranking prioritizes hazardous patterns, such as read-after-write dependencies across sessions. More generally, we score traces against hazard templates: short, parameterized overlap patterns defined by an anchor event (e.g., a read or commit boundary), one or more conflicting events on the same logical entity, and an optional terminal event that makes the effect observable, all within a bounded event window.

\textbf{Targeted amplification:} Finally, the selected ensemble undergoes amplification on the live SUT. When a failure is detected, the framework narrows the stochastic search window around the failing interleaving and re-executes the selected candidate under controlled scheduling variation and fixed pseudorandom seeds. The fixed seeds preserve the generated schedule decisions, while repeated execution continues to exercise the timing-sensitive region of the live SUT. This increases the probability of reproducing failure-prone interleavings and produces replayable regression candidates.

A lightweight test-side context links syntactically valid steps into semantically valid executions by tracking created entities, expected state, and cross-story preconditions. Event effects update this context as execution progresses, while bindings query it to select valid entities and preserve relationships across dependent operations. In the Magento model, for example, context updates record cart effects across \texttt{AddToCart} and \texttt{Checkout} events so that concurrent stories can verify the cart state observed during checkout. This mechanism supports both stateful API prefixes and multi-session workflow checking without requiring the stories themselves to encode low-level execution state.

\subsection{Domain-Independent Hazard Templates}
Our prioritization workflow derives lightweight indicators from recurring architectural risks. Hazard templates capture three general patterns:
\begin{itemize}
    \item \textbf{Multi-session resource contention:} Prioritize read-after-write or write-after-read orderings across overlapping actor sessions that access the same logical resource.
    \item \textbf{Semantic-prefix depth:} Prioritize compositions that construct and preserve long, syntactically valid state paths before a terminal validation call.
    \item \textbf{Business-rule boundaries:} Prioritize repeated or concurrent operations near cardinality, quota, exclusivity, and workflow-transition boundaries.
\end{itemize}

\subsection{Use of Generative AI Tools}
During parts of the research lifecycle, we used OpenAI's ChatGPT and other generative AI tools as engineering and verification aids. They assisted with drafting and debugging experiment-orchestration, evidence-collection, data-processing, consistency-checking, and artifact-packaging scripts; validating replication and evidence audits; cross-checking calculations and reported summaries; and editing and restructuring the manuscript.

\section{Evaluation}\label{sec:evaluation}
The evaluation examines two complementary failure dimensions and then synthesizes their common mechanism. ~\autoref{sec:magento-case} addresses RQ1 by evaluating ordering sensitivity and interleaving-aware amplification on Magento. ~\autoref{sec:rest-case} addresses RQ2 through a controlled Basic-versus-Complex SBT comparison and external schema-driven baselines across four REST API systems. RQ3 integrates the mechanism evidence from both studies to determine how admissible story compositions preserve failure-relevant executions across the two dimensions. Generated-model scale and estimated preparation-and-validation effort are reported separately as feasibility evidence.

\subsection{Case Study I: Magento E-Commerce}\label{sec:magento-case}
\subsubsection{Overview}
We demonstrate the method on Magento, part of the Adobe Commerce product family~\cite{adobe2024magento}. Independent web-technology telemetry reported Adobe Commerce on 2.5\% of the top 1,000,000 websites whose content-management system was known on September 5, 2026~\cite{w3techs2026adobecommerce}, supporting its use here as a mature real-world evaluation target. Magento is suitable because it supports rich end-to-end user workflows across multiple architectural layers, including workflows whose outcomes may be sensitive to concurrent session overlap. Its official testing stack includes PHPUnit for backend logic, Jasmine for frontend behaviors, and end-to-end suites via the Magento Functional Testing Framework (MFTF) and Selenium~\cite{adobe2024mftf}. These practices primarily validate short, maintainable functional workflows; our evaluation targets deliberately interleaved business flows.

\subsubsection{Why Magento is a meaningful concurrency target}
Conventional functional regression suites commonly exercise workflows independently, while load testing targets performance and leaves application-specific semantic invariants under particular cross-session interleavings outside its primary scope. Public reports and patches document race conditions, deadlocks, and overlap-sensitive checkout failures in Magento deployments~\cite{stackoverflow2010race,adobe2025magento247,adobe2024cloudpatches,nist2024cve45120,cve202549558}. Known Magento business-logic races provide ground truth for failure attribution, while the sporadic Shipping Fee target enables estimation of reproduction probability. Magento therefore provides a realistic environment for two complementary interleaving tests: a controlled ordering intervention on Empty Cart and amplification of a distinct sporadic Shipping Fee race.

\subsubsection{Method Application and Validation}
We instantiated the testing workflow by defining the scope around the shopping and checkout processes.

\textbf{Modeling and binding:} We composed independent stories for add-to-cart and checkout activities using a high-level vocabulary (\texttt{Login}, \texttt{AddToCart}, \texttt{Checkout}). Constraints encoded genuine business preconditions, while the relative ordering between the concurrent add-to-cart and checkout flows remained available for evaluation. These events were bound to Selenium WebDriver actions, with a shared context maintaining the expected cart state for verification.

\textbf{Validation:} During model stabilization, the preserved 100-test sampled suite exposed the Empty Cart signature under a two-session execution in which checkout login preceded completion of the concurrent \texttt{AddToCart} story. Both business flows remained present. We then changed only their relative ordering by requiring \texttt{AddToCart} completion before checkout login and executed two subsequent 100-run campaigns under the stabilized model.

\textbf{Exploration and amplification:} We then generated a large pool of interleavings and selected ensembles based on hazardous overlap windows. We amplified the selected tests through repeated execution, increasing the reproduction probability of the sporadic race manifestation.

\subsubsection{Business-Logic Race Reproduction and Analysis}
\textbf{The ``Empty Cart'' Business-Logic Race (Issue \#37465)~\cite{magentoIssue37465}:} This failure manifested when checkout login preceded completion of a concurrent \texttt{AddToCart} story, so checkout did not observe the newly added cart state. Under the preserved hazardous ordering, the Empty Cart signature appeared in $100/100$ executions; after constraining only the relative ordering so that \texttt{AddToCart} completed before checkout login, it appeared in $0/200$ follow-up executions. Both business flows remained present throughout the intervention. These counts are derived directly from the preserved 100-test sampled suite and the three corresponding campaign logs. The result therefore provides evidence of schedule sensitivity, not a second amplification experiment, and does not attribute a particular implementation-level data-race root cause.

\textbf{The ``Shipping Fee'' Business-Logic Race (Issue \#37466)~\cite{magentoIssue37466}:} We use this issue as the target business-logic race. Its manifestation required a narrow shipping/checkout overlap in which an \texttt{AddToCart} event occurred for a specific user while that user was in the process of checking out. Standard random testing rarely hits this timing. However, by applying the method's amplification phase, which ranked test ensembles based on their proximity to this hazardous interleaving, the selected set substantially increased the probability of reproducing this rare race manifestation under repeated replay.


\begin{table}[b]
\centering
\caption{Two complementary interleaving experiments in the Magento case study. Empty Cart evaluates sensitivity to a controlled ordering intervention, whereas Shipping Fee evaluates overlap-aware amplification against random replay; the columns therefore represent different experimental roles rather than parallel effect-size estimates.}
\label{tab:magento-interleaving}
\small
\setlength{\tabcolsep}{2.4pt}
\renewcommand{\arraystretch}{1.06}
\begin{tabular}{@{}>{\raggedright\arraybackslash}p{0.25\linewidth}>{\raggedright\arraybackslash}p{0.31\linewidth}>{\raggedright\arraybackslash}p{0.38\linewidth}@{}}
\toprule
\textbf{Evidence} & \textbf{Empty Cart} & \textbf{Shipping Fee} \\
\midrule
Role & Ordering intervention & Overlap-aware amplification \\
Comparison & Hazardous vs. constrained ordering & Top-ranked vs. random control \\
Campaign design & 100-run baseline; then $2\times100$ constrained runs & 10,000 sampled; 14 ranked schedules; $64$ replays each; 1,000 random-control runs \\
Target outcome & $100/100 \rightarrow 0/200$ & $369/896$ vs. $36/1{,}000$ \\
Rate & $100\% \rightarrow 0\%$ & $41.2\%$ [$28.1\%$, $54.3\%$] vs. $3.6\%$ [$2.6\%$, $4.9\%$] \\
Interpretation & Schedule-sensitive manifestation; not amplification & Amplification: $+37.6$ pp; RR $=11.44$ \\
Preserved evidence & 100-test sampled suite + 3 campaign logs & 14 replay schedules + random-control campaign \\
\bottomrule
\end{tabular}
\end{table}

\subsubsection{Amplification Results}
We use uniform-random replay as the black-box amplification baseline because established concurrency heuristics target instrumented program threads rather than cross-session web-level schedules~\cite{eytani2003heuristics,krena2012noise,weiss2025black}. The ranking and ensemble-selection procedure reduced an offline sample of 10,000 admissible schedules to a final ensemble of 14 schedules with high estimated exposure to the Shipping Fee hazard. Each selected schedule was replayed 64 times, yielding 896 amplified runs, while 1,000 randomly sampled schedules estimated the unguided baseline probability.

The ranking logic scores tests by searching, within a sliding event window, for patterns in which checkout overlaps with add-to-cart success and a follow-up callback before the order number is emitted. We fixed the window size at 20 events after preliminary pilot exploration because it covered the local event distances observed around candidate shipping/checkout overlaps while remaining narrow enough to discriminate among schedules. Neighboring values in the 15--25 range served as a tuning sanity check. We therefore treat 20 as an operational ranking parameter and limit its interpretation to schedule ranking.

For Shipping Fee, the amplified set reproduced the target business-logic race condition in $41.2\%$ of runs (369/896), compared with $3.6\%$ in the 1,000-test random control ($36/1{,}000$), an aggregate increase of 37.6 percentage points and a risk ratio of 11.44 (see ~\autoref{tab:magento-interleaving}). Because the amplified campaign comprises 64 repetitions of each of 14 selected schedules, schedule identity, rather than contiguous replay-log position, is the relevant grouping unit for schedule-level variability. Grouping executions by preserved replay-input schedule identity yields a 95\% $t$-based interval of $28.1\%$--$54.3\%$ across the 14 schedule-level reproduction rates; the control's 95\% Wilson interval is $2.6\%$--$4.9\%$. The wider schedule-level interval reflects substantial heterogeneity among the selected schedules and avoids treating all 896 repeated replays as mutually independent.

\begin{figure}[t]
\centering
\begin{tikzpicture}
\begin{axis}[
    ybar stacked,
    bar width=7.5pt,
    width=0.92\linewidth,
    height=5.6cm,
    ymin=0,
    ymax=55,
    ytick={0,20,40},
    ylabel={failures (of 64)},
    xlabel={selected schedule},
    symbolic x coords={1,2,3,4,5,6,7,8,9,10,11,12,13,14},
    xtick=data,
    nodes near coords,
    every node near coord/.append style={font=\footnotesize},
    legend style={at={(0.5,1.20)}, anchor=north, legend columns=-1, font=\small},
    legend entries={Target bug, Other failures},
    area legend
]
    \addplot+[
        ybar,
        fill=blue!35,
        draw=blue,
        nodes near coords style={text=blue, anchor=south, yshift=4pt}
    ] coordinates {
        (1,32) (2,38) (3,2) (4,39) (5,25) (6,43) (7,42) (8,29) (9,30) (10,33) (11,31) (12,4) (13,21) (14,0)
    };

    \addplot+[
        ybar,
        fill=red!35,
        draw=red,
        nodes near coords style={text=red, anchor=south, yshift=-1pt}
    ] coordinates {
        (1,1) (2,10) (3,17) (4,7) (5,19) (6,7) (7,0) (8,8) (9,6) (10,8) (11,4) (12,7) (13,15) (14,3)
    };
\end{axis}
\end{tikzpicture}
\caption{Per-schedule outcomes across the $14 \times 64$ Magento replay campaign, grouped by preserved replay-input schedule identity. Blue bars show Shipping Fee target reproductions, while red bars show other recorded failures from the same 64 replays; the latter are reported separately and are not included in the target-failure count.}
\label{fig:magento_chart}
\Description{Stacked bar chart for 14 selected Magento schedules, each replayed 64 times. Target-failure counts vary substantially across schedules, from zero or a few reproductions for some schedules to more than 40 for others; smaller stacked segments show non-target recorded failures.}
\end{figure}
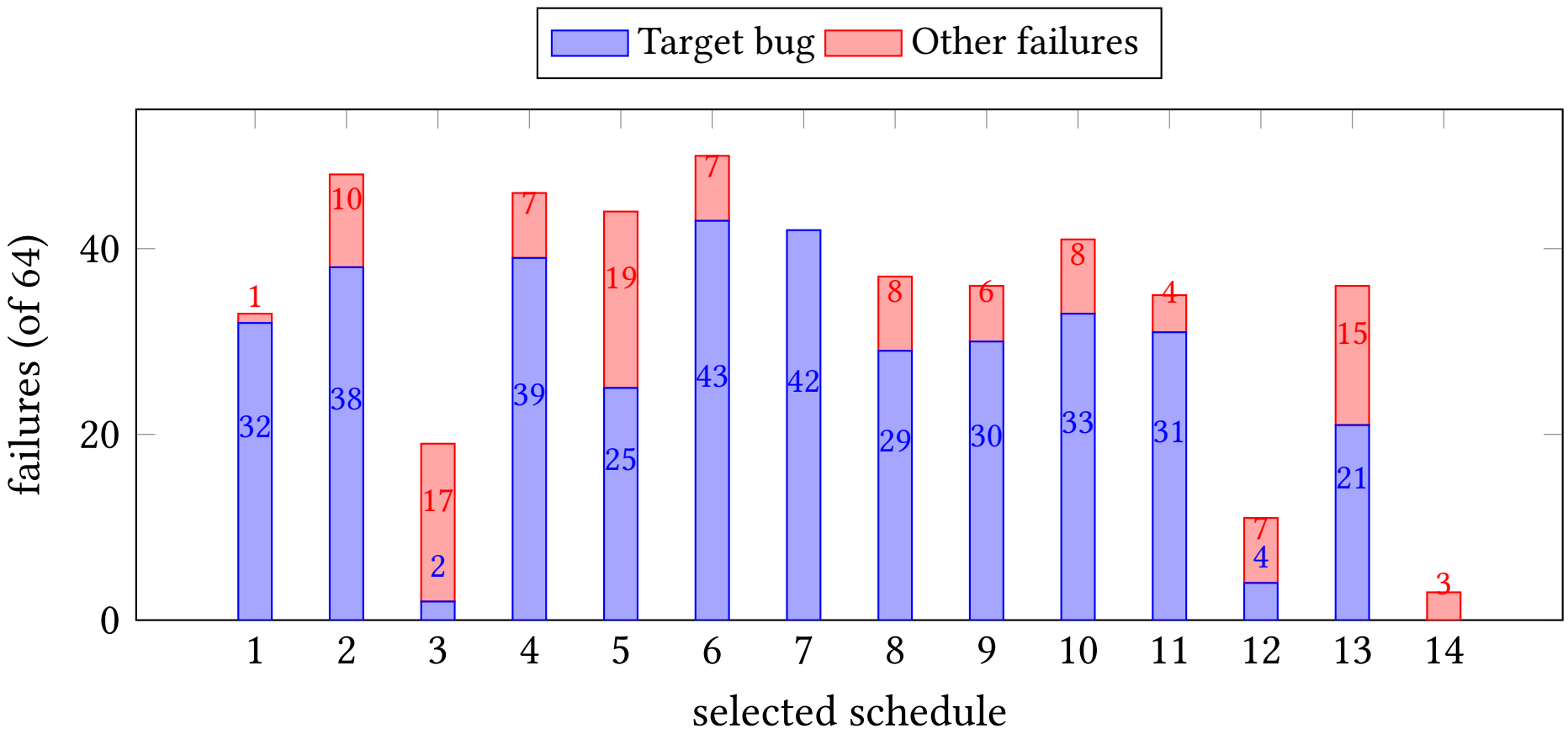

~\autoref{fig:magento_chart} shows this heterogeneity directly. Several selected schedules reproduced the target frequently, while others did so rarely or not at all; in aggregate, however, the selected set reproduced the target substantially more often than the unguided control. This result supports the overlap-aware ranking and repeated-replay stage described in~\autoref{sec:method} without implying that every highly ranked schedule has the same reproduction probability.

\paragraph{Answer to RQ1 (Interleaving).} The Empty Cart intervention establishes schedule sensitivity: preserving both business flows while constraining only their relative ordering eliminated the observed manifestation in the two follow-up campaigns. For the distinct sporadic Shipping Fee target, overlap-aware ranking increased aggregate reproduction from $3.6\%$ to $41.2\%$ (RR $=11.44$) relative to unguided replay; the 14 selected schedules had a schedule-level 95\% $t$-based interval of $28.1\%$--$54.3\%$. Thus, the Magento study provides evidence for both controlled ordering and targeted amplification, while the quantitative amplification estimate remains specific to Shipping Fee and to the selected high-risk schedules.

\subsection{Case Study II: Stateful REST APIs}\label{sec:rest-case}
Case Study II evaluates the depth axis: whether a testing method can preserve the legal, state-building prefix required before a business-rule violation becomes reachable. We operationalize a like-for-like comparison through two nested experimental contracts: a controlled SBT contrast that isolates generated scenario composition, and a cross-tool contract that evaluates complete testing methods at the same observable black-box boundary.

\subsubsection{Comparison Contract and Automatically Generated Configurations}
\label{sec:uc2-contract}

\paragraph{Controlled SBT contract.}
The versioned OpenAPI-to-SBT pipeline automatically transforms each frozen OpenAPI contract into executable Basic and Complex configurations, including stories, dependency relations, and interface bindings. The contract is the sole benchmark-specific generation input. The generator, runtime, SUT, clean-state policy, HTTP boundary, evaluator, and execution protocol remain fixed; only the declared generation profile changes. Generation uses no analyst-authored scenarios, fault identifiers, target traces, privileged SUT state, or evaluator feedback.

Both profiles cover all 86/86 operations and share the same contract-derived dependencies, bindings, and evaluator. In Library, Garage, and Pharmacy, Basic uses 1/1 entity/action multiplicities and Complex uses 5/7. In NetBox, both use 5/7; Basic performs direct generation, whereas Complex enables dependency-derived composition. The contrast therefore measures the reachability gained from automated composition richness: multiplicity and composition in the compact benchmarks, and composition alone in NetBox.

\paragraph{Cross-tool contract.}
Basic SBT, Complex SBT, RESTler, and EvoMaster are evaluated as end-to-end tool configurations on the same externally observable task. All four configurations receive the same frozen public OpenAPI and task-relevant operation set, execute the same SUT through the same HTTP boundary under the same clean-state policy, and are scored against the same semantic-class ledger by the same external evaluator. Each method transforms the common public input through its native generation procedure, while the task, evidence unit, and confirmation rule remain fixed. This contract measures end-to-end semantic reachability under common information and observation conditions, consistently with testing evaluations that preserve native tool procedures~\cite{beyer2021testcomp,martinlopez2020restest,karlsson2024behaviours}.

\paragraph{Official evidence contract.}
The official evidence unit is one independently delimited test generated by one evaluated configuration. Before replaying each test, the SUT is restored to the declared clean state. A semantic class is confirmed only when that test alone establishes both the complete legal prefix and the class-specific invariant violation. Evidence is never joined across generated tests or resets. Campaign-global traces report diagnostic reachability only; native tool labels, schema mismatches, timeouts, and generic HTTP failures do not determine the semantic score. Missing test boundaries or incomplete replay produce an undetermined result rather than an inferred zero. Each predeclared semantic class is counted at most once per configuration.

\subsubsection{Benchmark Construction Protocol}
Library, Garage, and Pharmacy are controlled benchmarks with auditable ground truth covering lending and reservation invariants, shared-resource coordination, exclusivity, and regulated workflows. The seeded faults are absent from every tool's generation input. SBT automatically derives and executes its stories, dependencies, and bindings from the frozen OpenAPI contract; RESTler and EvoMaster apply their native automatic pipelines. The common semantic evaluator is applied only to completed executions. The artifact retains the SUTs, contracts, generated models, reports, and fault definitions.

\begin{lstlisting}[language=JavaScript, caption={Auto-generated high-level Pharmacy story. The story requests the logical lifecycle of an order without exposing HTTP details.}, label={lst:story}]
bthread("crud:Orders:linear:1", function () {
  let id = "id_260_" + Math.floor(Math.random() * 10000);
  createOrder(id);
  verifyOrdersExists(id);
  updateOrder(id);
  deleteOrder(id);
  tryToDeleteANonExistingOrders(id);
  verifyOrdersDoesNotExist(id);
});
\end{lstlisting}

\begin{lstlisting}[language=JavaScript, caption={Auto-generated Pharmacy interface binding. The binding maps one logical story step to a concrete \texttt{POST /orders} request and emits a completion event to the BP model.}, label={lst:binding}]
function createOrder(id) {
  var url = "/orders";
  var description = "Create order " + id;
  var body = { "id": String(id) };
  svc.post(url, {
    body: JSON.stringify(body),
    expectedResponseCodes: [201, 400, 409],
    parameters: { description: description }
  });
  bp.sync({ request: bp.Event("Done: " + description, { "id": id }) });
}
\end{lstlisting}

~\autoref{lst:story} and ~\autoref{lst:binding} show the generated high-level/low-level split: the story expresses business intent, and the binding translates a logical step into an HTTP request. Magento uses the same separation with analyst-authored Selenium bindings.

The important point is that this story does not execute in isolation. In the Pharmacy benchmark, ordinary CRUD stories compose with automatically generated dependency-aware flows over refill and order-processing operations. BP therefore explores admissible compositions among different stories: one story may create or update an order, another may establish a prescription and store, and a third may issue a process or refill action. A confirming execution can thus emerge from the generated composition rather than from a benchmark-specific linear script.

Each compact synthetic system includes two auditable semantic fault classes. The artifact retains the benchmark SUTs, executable story models, and class-specific runtime evidence, so the reported counts are based on observed executions rather than enabled defect flags. In Library (8 paths, 13 operations), Loan Overflow allows a third active loan after two legal loans, and Hold Theft allows a second user to borrow a book already held by another user. In Garage (14 paths, 32 operations), Garage Overflow opens a third active repair order in the same garage, and Teleporting Car assigns the same car to simultaneous active orders in different garages. In Pharmacy (10 paths, 29 operations), Refill Underflow allows repeated dispenses to drive \texttt{refillsLeft} below zero, and Double Process lets a prescription already locked by one store be processed again by another. These benchmarks are compact, but they represent recurring stateful business-rule failures: cardinality constraints, reservation conflicts, shared-resource exclusivity, regulated workflow violations, and duplicate processing.

Two supplementary Library controls probe depth without changing the primary 11-class ledger. The short-prefix control reuses the two primary Library classes with shallower setup. The extended-prefix control introduces a separate false-rejection class: after four distinct loans have each been returned, the public active-loan count is zero and a fifth distinct loan is legal, but stale SUT state rejects it as if the active limit had been exceeded. Its public contract explicitly states the limit and that returned loans are inactive.

NetBox contributes a large real-world API specification drawn from an organically evolved open-source system. We use the full public specification to assess modeling scale. Separately, for controlled and auditable fault attribution, the semantic-fault experiment uses the NetBox projection defined above against a benchmark harness (\path{netbox_benchmark.py}) implementing five predeclared semantic fault classes. They span state/precondition, lifecycle, uniqueness, integrity, and workflow violations. The harness preserves the projected API inputs and emits execution evidence from which class-specific oracles confirm the legal prefix and invariant violation.

~\autoref{tab:results} counts each of the 11 predeclared semantic fault classes once under one evaluation rule and one evidence schema across all four systems and all four configurations. A class is confirmed only when one retained sequence establishes both its legal semantic prefix and its class-specific invariant violation. The final ledger resolves every configuration-by-class cell from runtime evidence; no enabled flag, generic server error, timeout, schema mismatch, or campaign-global event stitching is counted as a semantic confirmation.

The final evidence release applies the same class-level decision rule to every retained primary execution. It contains outcomes for the 44 configuration-by-class cells (11 classes $\times$ four configurations), together with the generator inputs and reports, executable models, contracts, SUTs, logs, generated tests, replay boundaries, and class-specific semantic evaluations. Basic SBT confirmed 0/11 classes, Complex SBT confirmed 11/11, RESTler confirmed 2/11, and EvoMaster confirmed 0/11. RESTler's two confirmations are the NetBox uniqueness and referential-integrity classes; every reported value comes from the final canonical sequence-local ledger.

\paragraph{Execution protocol and campaign scale.}
All Case Study II runs were executed on a machine equipped with a 12th Gen Intel Core i5-1235U processor (10 physical cores, 12 logical processors) and 15.6~GiB of RAM, running 64-bit Microsoft Windows 11 Enterprise. RESTler and EvoMaster were each configured with a nominal one-hour search budget on each REST benchmark. Preserved wall-clock durations also include startup, checker execution, shutdown, and post-processing; RESTler runs can therefore exceed \texttt{--time\_budget 1.0}. The retained Garage EvoMaster run executed 54,681 actions. SBT is completion-bounded: the retained primary Library campaign completed in 153.5~s. For the extended-prefix control, we selected one composition of the contract-derived stories to isolate whether SBT could preserve and execute the required four-cycle semantic prefix. Its 4.924-s duration documents reachability; the baselines' one-hour budgets characterize autonomous search. The artifact preserves all configurations, stopping criteria, and runtime evidence.

To characterize generated-model scale and adoption cost, ~\autoref{tab:models} reports retained model structure and retrospective preparation-and-validation effort across the evaluated systems. The reported effort ranges from 1.5--2.5 person-hours for the compact APIs to 12--15 person-hours for NetBox. The structural counts are directly reproducible, while the effort estimates characterize engineering feasibility. For Magento, which is modeled rather than generated from OpenAPI, the table reports the full authored model; the RQ1-specific shopping-and-checkout slice uses three story units, including one anti-story, and four event definitions, three of which perform Selenium interface actions and one of which serves as a synchronization marker.

\begin{table*}[t]
\centering
\caption{Structural sizes and estimated preparation-and-validation effort for the executable models. Story units count generated or authored story/b-thread declarations, anti-stories are the \texttt{block(...)} subset, and bindings connect story-level events to interface behavior or synchronization semantics. Effort values are retrospective engineering estimates rather than instrumented development-time measurements.}
\label{tab:models}
\small
\resizebox{0.98\textwidth}{!}{%
\begin{tabular}{lccccc}
\toprule
\textbf{System} & \shortstack{\textbf{Story}\\\textbf{Units}} & \textbf{Anti-stories} & \textbf{Bindings} & \shortstack{\textbf{Estimated}\\\textbf{Preparation/Validation}\\\textbf{(person-hours)}} & \textbf{Notes} \\
\midrule
Magento  & 6   & 1  & 7     & $\sim$4--6     & Selenium UI + administrator workflow. \\
Library  & 14  & 0  & 29    & $\sim$1.5--2.0 & CRUD + in-memory state tracker. \\
Garage   & 25  & 1  & 77    & $\sim$2.0--2.5 & Shared-resource coordination. \\
Pharmacy & 22  & 3  & 63    & $\sim$2.0--2.5 & Multi-step workflow invariants. \\
NetBox   & 453 & 61 & 1,514 & $\sim$12--15   & Full spec; faults use 12-operation projection. \\
\bottomrule
\end{tabular}
}
\end{table*}

\subsubsection{Results and Analysis}
~\autoref{tab:results} reports the controlled SBT contrast and the end-to-end comparison with external baselines across 11 unique semantic fault classes. ~\autoref{tab:depth-controls} separately reports the supplementary Library depth controls; these results support mechanism analysis and are excluded from the primary total. Under the common sequence-local confirmation rule, Basic SBT confirmed 0/11 classes, Complex SBT confirmed 11/11, RESTler confirmed 2/11, and EvoMaster confirmed 0/11.

\begin{table*}[t]
\centering
\begin{minipage}[t]{0.61\textwidth}
\centering
\captionsetup{margin=4pt}
\captionof{table}{Official sequence-local confirmations under the controlled SBT and cross-tool contracts (confirmed/eligible classes). Basic and Complex are automatically generated profiles of the same SBT pipeline; RESTler and EvoMaster are end-to-end external baselines.}
\label{tab:results}
\scriptsize
\setlength{\tabcolsep}{1.8pt}
\begin{tabular}{@{}lcccc@{}}
\toprule
& \multicolumn{2}{c}{\textbf{Controlled SBT}} & \multicolumn{2}{c}{\textbf{External}} \\
\cmidrule(lr){2-3}\cmidrule(l){4-5}
\textbf{System} & \textbf{Basic} & \textbf{Complex} & \textbf{RESTler} & \textbf{EvoMaster} \\
\midrule
Library  & 0/2 & 2/2 & 0/2 & 0/2 \\
Garage   & 0/2 & 2/2 & 0/2 & 0/2 \\
Pharmacy & 0/2 & 2/2 & 0/2 & 0/2 \\
NetBox   & 0/5 & 5/5 & 2/5 & 0/5 \\
\midrule
\textbf{Total} & \textbf{0/11} & \textbf{11/11} & \textbf{2/11} & \textbf{0/11} \\
\bottomrule
\end{tabular}
\end{minipage}
\hfill
\begin{minipage}[t]{0.37\textwidth}
\centering
\captionsetup{margin=4pt}
\captionof{table}{Supplementary Library depth controls, excluded from the primary total. Short prefix reuses two primary classes; extended prefix is a separate control class.}
\label{tab:depth-controls}
\scriptsize
\setlength{\tabcolsep}{2.0pt}
\resizebox{\linewidth}{!}{%
\begin{tabular}{@{}lccc@{}}
\toprule
\textbf{Prefix} &
\textbf{RESTler} &
\textbf{EvoMaster} &
\shortstack{\textbf{Complex}\\\textbf{SBT}} \\
\midrule
Short    & 2/2 & 0/2 & 2/2 \\
Extended & 0/1 & 0/1 & 1/1 \\
\bottomrule
\end{tabular}
}
\end{minipage}
\vspace{-8pt}
\end{table*}

RESTler and EvoMaster also reported interface- and contract-level outcomes, such as timeouts, invalid-value responses, and schema or status mismatches. We retain these tool-reported outcomes in the artifact, but do not label them as distinct bugs or include them in the semantic counts because they do not satisfy the confirmation rule defined in ~\autoref{sec:oracles-static}.

The controlled profile intervention changed the confirmed total from 0/11 to 11/11 while holding the OpenAPI input, generator, bindings, runtime, SUT, and evaluator fixed. In the three compact benchmarks, Complex increases the multiplicities from 1/1 to 5/7 and enables dependency-derived compositions; in NetBox, both profiles retain 5/7 and differ only in composition. The result attributes the gain to the automatically constructed, dependency-preserving composition space.

The supplementary Library evidence identifies the reachability mechanism more closely. The short-prefix control shows that RESTler confirms both Library classes when their required prefixes are shallow, while EvoMaster confirms neither; it also establishes that the endpoints and common evaluator were operational. In the extended-prefix control, the selected Complex SBT composition executed four successful loan/return cycles for distinct books and then confirmed an erroneous \texttt{409} rejection despite zero active loans in 4.924~s. Against the same SUT, RESTler completed five successful loans and one return but never exceeded one returned distinct book for a user, below the four required. EvoMaster evaluated 15,538 tests (49,717 actions) in one hour and successfully exercised nine of 13 endpoints, including user and book creation, but its generated \texttt{POST /loans} requests returned \texttt{422} responses for unknown references because their identifiers were not bound to resources created within the same test. Thus, neither baseline constructed the required four-cycle prefix, whereas the Complex profile realized it end to end. These retained runs establish depth-sensitive reachability under the declared execution contract.

\paragraph{Answer to RQ2 (Depth).} Automatic, parameterized scenario construction determined depth-sensitive reachability. From the same frozen OpenAPI and generation pipeline, Basic and Complex confirmed $0/11$ and $11/11$ semantic classes. The binding ablation ($0/250$ versus $250/250$ valid dependent operations) and Library prefix controls localize the mechanism to preservation of concrete relations and legal state-building prefixes. Under the shared cross-tool contract, RESTler and EvoMaster confirmed $2/11$ and $0/11$ classes. Thus, generated composition richness changed semantic reachability without changing the public input, SUT boundary, or sequence-local evidence rule.

\paragraph{Answer to RQ3 (Cross-Axis Mechanism).} Across both axes, an admissible story composition is the common search object. On the depth axis, automatic composition richness and semantic binding preserve legal state-building prefixes. On the interleaving axis, event eligibility generates alternative legal schedules, while ranking and replay select and amplify failure-relevant overlaps. The Basic--Complex contrast, binding ablation, Empty Cart ordering intervention, and Shipping Fee amplification consistently support these mechanisms. Thus, semantic constraints preserve business validity while composition structure determines which deep states and cross-session schedules become reachable; this connects the two axes without requiring identical search procedures.

\FloatBarrier

\section{Threats to Validity}\label{sec:threats}

\textbf{Internal validity.} All 11 primary REST fault classes and four configurations use the same class-specific rule: one sequence replayed from a clean SUT must establish a legal semantic prefix and its invariant violation. Timeouts, invalid values, schema/status mismatches, generic 500-class outcomes, campaign-global stitching, and incomplete evidence are not confirmations. Magento target and non-target failures are counted separately; Empty Cart additionally requires the preserved ordering and class-specific signature.

\textbf{Construct validity.} The three compact REST benchmarks contain seeded rather than historically mined defects and represent targeted families, not defect prevalence. Basic and Complex use the same frozen OpenAPI and versioned pipeline, making the generation profile the controlled variable. The profile changes multiplicity and dependency-derived composition in the three compact systems, but only dependency-derived composition in NetBox, where both profiles retain 5/7. We mitigate possible shared generator defects by retaining rules, parameters, provenance, generated artifacts, and hashes, and auditing operation and dependency coverage. Across tools, the public input, SUT, clean-state policy, semantic-class ledger, and sequence-local evaluator are shared, while each method retains its native automatic transformation. The evaluator and fault definitions are external to generation. The ablation and prefix controls establish mechanism and reachability; effort estimates characterize engineering feasibility.

\textbf{External validity.} Library, Garage, and Pharmacy are author-constructed benchmarks with seeded defects that isolate recurring state-dependent mechanisms. NetBox assesses modeling scale from its full public specification; fault attribution uses a task-relevant 12-operation projection and benchmark SUT. Empty Cart establishes schedule sensitivity, while Shipping Fee supplies the quantitative amplification evidence. These choices improve attribution but limit generalization to organically evolved faults and other systems.

\textbf{Conclusion validity.} Shipping Fee uncertainty is summarized over 14 schedule identities, while the random control uses a Wilson interval. Deliberate selection quantifies amplification within the high-risk set. Empty Cart establishes ordering sensitivity, and Shipping Fee supplies the quantitative amplification estimate.

\section{Conclusion and Future Work}\label{sec:conclusion}

Story-based semantic fuzzing uses admissible story compositions to search deep states and cross-session interleavings. On Magento, overlap-aware ranking increased Shipping Fee reproduction from $3.6\%$ to $41.2\%$, while the Empty Cart intervention established ordering sensitivity. Across four REST APIs, automatically generated Basic and Complex profiles confirmed $0/11$ and $11/11$ semantic classes; RESTler and EvoMaster confirmed $2/11$ and $0/11$ under the shared sequence-local contract. Prefix controls and the $250/250$ versus $0/250$ binding ablation localize the mechanism to preserving legal relational state. Together, the studies show how semantic composition supports both depth and interleaving-sensitive testing.

\textbf{Future work} will extend generator conformance audits, develop bug-independent objectives for deep semantic prefixes, and evaluate further industrial systems while retaining analyst validation where contracts are ambiguous.

\paragraph{Data Availability.} An anonymized supplementary replication package contains the retained evidence, executable models, semantic evaluators, and scripts needed to regenerate and verify Tables~2--5, figures, and reported statistics. Author-created source code and scripts are released under the MIT License; author-created documentation, research data, and retained experimental evidence are released under the Creative Commons Attribution 4.0 International License (CC BY 4.0). Included third-party materials retain their original licenses.


\bibliographystyle{ACM-Reference-Format}
\bibliography{references}

@misc{adobe2024magento,
  author       = {{Adobe}},
  title        = {{Magento} transitions to {Adobe Commerce}},
  year         = {2025},
  url          = {https://business.adobe.com/products/commerce/magento.html},
  note         = {Accessed September 3, 2026}
}

@misc{adobe2024cloudpatches,
  author       = {{Adobe}},
  title        = {Cloud Patches for Commerce},
  year         = {2026},
  url          = {https://experienceleague.adobe.com/en/docs/commerce-on-cloud/user-guide/release-notes/cloud-patches},
  note         = {Last updated August 19, 2026. Accessed September 3, 2026}
}

@misc{adobe2025magento247,
  author       = {{Adobe Commerce}},
  title        = {{Magento Open Source} 2.4.7 release notes},
  year         = {2026},
  url          = {https://experienceleague.adobe.com/en/docs/commerce-operations/release/notes/magento-open-source/2-4-7},
  note         = {Last updated August 19, 2026. Accessed September 3, 2026}
}

@misc{adobe2024mftf,
  author       = {{Adobe Commerce}},
  title        = {Getting started with functional testing},
  year         = {n.d},
  url          = {https://developer.adobe.com/commerce/testing/functional-testing-framework/getting-started},
  note         = {Accessed September 3, 2026}
}

@article{arcuri2025tool,
  author       = {Andrea Arcuri and Man Zhang and Susruthan Seran and Juan Pablo Galeotti and Amid Golmohammadi and Onur Duman and Agustina Aldasoro and Hernan Ghianni},
  title        = {Tool report: {EvoMaster}---black and white box search-based fuzzing for {REST}, {GraphQL} and {RPC} APIs},
  journal      = {Automated Software Engineering},
  volume       = {32},
  number       = {1},
  articleno    = {4},
  numpages     = {11},
  year         = {2025},
  doi          = {10.1007/s10515-024-00478-1}
}

@inproceedings{atlidakis2019restler,
  author       = {Vaggelis Atlidakis and Patrice Godefroid and Marina Polishchuk},
  title        = {{RESTler}: Stateful {REST API} Fuzzing},
  booktitle    = {Proceedings of the 41st International Conference on Software Engineering (ICSE '19)},
  pages        = {748--758},
  year         = {2019},
  publisher    = {IEEE},
  address      = {Piscataway, NJ, USA},
  doi          = {10.1109/ICSE.2019.00083}
}

@inproceedings{barsinai2019scenario,
  author       = {Michael Bar-Sinai and Achiya Elyasaf and Aviran Sadon and Gera Weiss},
  title        = {A Scenario Based On-Board Software and Testing Environment for Satellites},
  booktitle    = {Proceedings of the 59th Israel Annual Conference on Aerospace Sciences (IACAS '19)},
  volume       = {2},
  pages        = {1407--1419},
  year         = {2019},
  publisher    = {Israel Annual Conference on Aerospace Sciences},
  address      = {Tel-Aviv and Haifa, Israel},
}

@inproceedings{barsinai2023provengo,
  author       = {Michael Bar-Sinai and Achiya Elyasaf and Gera Weiss and Yeshayahu Weiss},
  title        = {{Provengo}: A Tool Suite for Scenario Driven Model-Based Testing},
  booktitle    = {Proceedings of the 38th IEEE/ACM International Conference on Automated Software Engineering (ASE '23)},
  pages        = {2062--2065},
  year         = {2023},
  publisher    = {IEEE},
  address      = {Piscataway, NJ, USA},
  doi          = {10.1109/ASE56229.2023.00146}
}

@inproceedings{barsinai2018bpjs,
  author       = {Michael Bar-Sinai and Gera Weiss and Reut Shmuel},
  title        = {{BPjs}: An Extensible, Open Infrastructure for Behavioral Programming Research},
  booktitle    = {Proceedings of the 21st ACM/IEEE International Conference on Model Driven Engineering Languages and Systems: Companion Proceedings (MODELS '18)},
  pages        = {59--60},
  year         = {2018},
  publisher    = {ACM},
  address      = {New York, NY, USA},
  doi          = {10.1145/3270112.3270126}
}

@inproceedings{bochmann1977unified,
  author       = {Gregor V. Bochmann and Jan Gecsei},
  title        = {A Unified Method for the Specification and Verification of Protocols},
  booktitle    = {Information Processing 77: Proceedings of the IFIP Congress},
  pages        = {229--234},
  year         = {1977},
  publisher    = {North-Holland Publishing Company},
  address      = {Amsterdam, The Netherlands}
}

@article{cao2025mixed,
  author       = {Shaoheng Cao and Renyi Chen and Wenhua Yang and Minxue Pan and Xuandong Li},
  title        = {A Mixed-Methods Study of Model-Based {GUI} Testing in Real-World Industrial Settings},
  journal      = {Proceedings of the ACM on Software Engineering},
  volume       = {2},
  number       = {FSE},
  pages        = {1545--1566},
  year         = {2025},
  doi          = {10.1145/3715789}
}

@incollection{clarke2012model,
  author       = {Edmund M. Clarke and William Klieber and Milo{\v{s}} Nov{\'a}{\v{c}}ek and Paolo Zuliani},
  title        = {Model Checking and the State Explosion Problem},
  booktitle    = {Tools for Practical Software Verification},
  series       = {Lecture Notes in Computer Science},
  volume       = {7682},
  pages        = {1--30},
  year         = {2012},
  publisher    = {Springer},
  address      = {Berlin, Heidelberg},
  doi          = {10.1007/978-3-642-35746-6_1}
}

@misc{cve202549558,
  author       = {{CVE Program}},
  title        = {{CVE-2025-49558}},
  year         = {2025},
  url          = {https://www.cve.org/CVERecord?id=CVE-2025-49558},
  note         = {CVE record. Accessed September 3, 2026}
}

@article{danglot2019automatic,
  author       = {Benjamin Danglot and Oscar Luis Vera-P{\'e}rez and Benoit Baudry and Martin Monperrus},
  title        = {Automatic Test Improvement with {DSpot}: A Study with Ten Mature Open-Source Projects},
  journal      = {Empirical Software Engineering},
  volume       = {24},
  number       = {4},
  pages        = {2603--2635},
  year         = {2019},
  doi          = {10.1007/s10664-019-09692-y}
}

@misc{stackoverflow2010race,
  author       = {Dyip302},
  title        = {Avoid race conditions in eCommerce scenarios},
  year         = {2010},
  howpublished = {Stack Overflow question 3389000},
  url          = {https://stackoverflow.com/questions/3389000/avoid-race-conditions-in-ecommerce-scenarios},
  note         = {Asked August 2, 2010. Accessed September 3, 2026}
}

@article{elyasaf2021context,
  author       = {Achiya Elyasaf},
  title        = {Context-Oriented Behavioral Programming},
  journal      = {Information and Software Technology},
  volume       = {133},
  pages        = {106504},
  year         = {2021},
  doi          = {10.1016/j.infsof.2020.106504}
}

@inproceedings{eytani2003heuristics,
  author       = {Yaniv Eytani and Eitan Farchi and Yosi Ben-Asher},
  title        = {Heuristics for Finding Concurrent Bugs},
  booktitle    = {Proceedings of the International Parallel and Distributed Processing Symposium (IPDPS '03)},
  pages        = {8},
  year         = {2003},
  publisher    = {IEEE},
  address      = {Piscataway, NJ, USA},
  doi          = {10.1109/IPDPS.2003.1213514}
}

@inproceedings{fonseca2010study,
  author       = {Pedro Fonseca and Cheng Li and Vishal Singhal and Rodrigo Rodrigues},
  title        = {A Study of the Internal and External Effects of Concurrency Bugs},
  booktitle    = {Proceedings of the 2010 IEEE/IFIP International Conference on Dependable Systems and Networks (DSN '10)},
  pages        = {221--230},
  year         = {2010},
  publisher    = {IEEE},
  address      = {Piscataway, NJ, USA},
  doi          = {10.1109/DSN.2010.5544315}
}

@article{formica2023search,
  author       = {Federico Formica and Tony Fan and Claudio Menghi},
  title        = {Search-Based Software Testing Driven by Automatically Generated and Manually Defined Fitness Functions},
  journal      = {ACM Transactions on Software Engineering and Methodology},
  volume       = {33},
  number       = {2},
  articleno    = {40},
  numpages     = {37},
  month        = dec,
  year         = {2023},
  doi          = {10.1145/3624745}
}

@article{garousi2021model,
  author       = {Vahid Garousi and Alper Bu{\u{g}}ra Kele{\c{s}} and Yunus Balaman and Zeynep {\"O}zdemir G{\"u}ler and Andrea Arcuri},
  title        = {Model-Based Testing in Practice: An Experience Report from the Web Applications Domain},
  journal      = {Journal of Systems and Software},
  volume       = {180},
  pages        = {111032},
  year         = {2021},
  doi          = {10.1016/j.jss.2021.111032}
}

@article{getto2020story,
  author       = {Guiseppe Getto},
  title        = {The Story/Test/Story Method: A Combined Approach to Usability Testing and Contextual Inquiry},
  journal      = {Computers and Composition},
  volume       = {55},
  pages        = {102548},
  year         = {2020},
  doi          = {10.1016/j.compcom.2020.102548}
}

@article{harel1987statecharts,
  author       = {David Harel},
  title        = {Statecharts: A Visual Formalism for Complex Systems},
  journal      = {Science of Computer Programming},
  volume       = {8},
  number       = {3},
  pages        = {231--274},
  year         = {1987},
  doi          = {10.1016/0167-6423(87)90035-9}
}

@inproceedings{harel2011model,
  author       = {David Harel and Robby Lampert and Assaf Marron and Gera Weiss},
  title        = {Model-Checking Behavioral Programs},
  booktitle    = {Proceedings of the Ninth ACM International Conference on Embedded Software (EMSOFT '11)},
  pages        = {279--288},
  year         = {2011},
  publisher    = {ACM},
  address      = {New York, NY, USA},
  doi          = {10.1145/2038642.2038686}
}

@incollection{harel2010programming,
  author       = {David Harel and Assaf Marron and Gera Weiss},
  title        = {Programming Coordinated Behavior in {Java}},
  booktitle    = {ECOOP 2010 -- Object-Oriented Programming},
  series       = {Lecture Notes in Computer Science},
  volume       = {6183},
  pages        = {250--274},
  year         = {2010},
  publisher    = {Springer},
  address      = {Berlin, Heidelberg},
  doi          = {10.1007/978-3-642-14107-2_12}
}

@article{harel2012behavioral,
  author       = {David Harel and Assaf Marron and Gera Weiss},
  title        = {Behavioral Programming},
  journal      = {Communications of the ACM},
  volume       = {55},
  number       = {7},
  pages        = {90--100},
  year         = {2012},
  doi          = {10.1145/2209249.2209270}
}

@inproceedings{hatfield2022deriving,
  author       = {Zac Hatfield-Dodds and Dmitry Dygalo},
  title        = {Deriving Semantics-Aware Fuzzers from Web {API} Schemas},
  booktitle    = {Proceedings of the 44th International Conference on Software Engineering: Companion Proceedings (ICSE-Companion '22)},
  pages        = {345--346},
  year         = {2022},
  publisher    = {ACM},
  address      = {New York, NY, USA},
  doi          = {10.1145/3510454.3528637}
}

@techreport{kaner2003introduction,
  author       = {Cem Kaner},
  title        = {An Introduction to Scenario Testing},
  institution  = {Florida Institute of Technology},
  address      = {Melbourne, FL},
  year         = {2003},
  url          = {https://kaner.com/pdfs/ScenarioIntroVer4.pdf},
  note         = {Accessed September 3, 2026}
}

@incollection{krena2012noise,
  author       = {Bohuslav K{\v{r}}ena and Zden{\v{e}}k Letko and Tom{\'a}{\v{s}} Vojnar},
  title        = {Noise Injection Heuristics for Concurrency Testing},
  booktitle    = {Mathematical and Engineering Methods in Computer Science},
  series       = {Lecture Notes in Computer Science},
  volume       = {7119},
  pages        = {123--135},
  year         = {2012},
  publisher    = {Springer},
  address      = {Berlin, Heidelberg},
  doi          = {10.1007/978-3-642-25929-6_12}
}

@inproceedings{lin2023forest,
  author       = {Jiaxian Lin and Tianyu Li and Yang Chen and Guangsheng Wei and Jiadong Lin and Sen Zhang and Hui Xu},
  title        = {{foREST}: A Tree-Based Black-Box Fuzzing Approach for {RESTful APIs}},
  booktitle    = {Proceedings of the 34th IEEE International Symposium on Software Reliability Engineering (ISSRE '23)},
  pages        = {695--705},
  year         = {2023},
  publisher    = {IEEE},
  address      = {Piscataway, NJ, USA},
  doi          = {10.1109/ISSRE59848.2023.00023}
}

@article{lonetti2023model,
  author       = {Francesca Lonetti and Antonia Bertolino and Felicita Di Giandomenico},
  title        = {Model-Based Security Testing in {IoT} Systems: A Rapid Review},
  journal      = {Information and Software Technology},
  volume       = {164},
  pages        = {107326},
  year         = {2023},
  doi          = {10.1016/j.infsof.2023.107326}
}

@inproceedings{lu2008learning,
  author       = {Shan Lu and Soyeon Park and Eunsoo Seo and Yuanyuan Zhou},
  title        = {Learning from Mistakes: A Comprehensive Study on Real World Concurrency Bug Characteristics},
  booktitle    = {Proceedings of the 13th International Conference on Architectural Support for Programming Languages and Operating Systems (ASPLOS '08)},
  pages        = {329--339},
  year         = {2008},
  publisher    = {ACM},
  address      = {New York, NY, USA},
  doi          = {10.1145/1346281.1346323}
}

@article{masud2017automated,
  author       = {Mahawish Masud and Muhammad Iqbal and M. U. Khan and Farooque Azam},
  title        = {Automated User Story Driven Approach for Web-Based Functional Testing},
  journal      = {International Journal of Computer and Information Engineering},
  volume       = {11},
  number       = {1},
  pages        = {91--98},
  year         = {2017},
  doi          = {10.5281/zenodo.1339932}
}

@book{meszaros2007xunit,
  author       = {Gerard Meszaros},
  title        = {xUnit Test Patterns: Refactoring Test Code},
  year         = {2007},
  publisher    = {Addison-Wesley Professional},
  address      = {Boston, MA}
}

@misc{nist2024cve45120,
  author       = {{National Institute of Standards and Technology}},
  title        = {{NVD}: {CVE-2024-45120}},
  year         = {2024},
  url          = {https://nvd.nist.gov/vuln/detail/CVE-2024-45120},
  note         = {National Vulnerability Database record. Accessed September 3, 2026}
}

@inproceedings{rooijakkers2025wuppiefuzz,
  author       = {Thomas Rooijakkers and Anne Nijsten and Cristian Daniele and Erieke Weitenberg and Ringo Groenewegen and Arthur Melissen},
  title        = {{WuppieFuzz}: Coverage-Guided, Stateful {REST API} Fuzzing},
  booktitle    = {Proceedings of the 12th International Conference on Information Systems Security and Privacy -- Volume 2: ICISSP},
  pages        = {221--231},
  year         = {2026},
  publisher    = {INSTICC, SciTePress},
  address      = {Set{\'u}bal, Portugal},
  doi          = {10.5220/0014327000004061}
}

@article{taromirad2025assertions,
  author       = {Masoumeh Taromirad and Per Runeson},
  title        = {Assertions in Software Testing: Survey, Landscape, and Trends},
  journal      = {International Journal on Software Tools for Technology Transfer},
  volume       = {27},
  pages        = {117--135},
  year         = {2025},
  doi          = {10.1007/s10009-025-00794-1}
}

@misc{weiss2021testing,
  author       = {Yeshayahu Weiss},
  title        = {Testing Reactive Systems Using Behavioural Programming, a Model Centric Approach},
  year         = {2021},
  archivePrefix= {arXiv},
  eprint       = {2112.01538},
  primaryClass = {cs.SE},
  doi          = {10.48550/arXiv.2112.01538}
}

@article{weiss2025black,
  author       = {Yeshayahu Weiss and Gal Amram and Achiya Elyasaf and Eitan Farchi and Oded Margalit and Gera Weiss},
  title        = {Black-Box Bug Amplification for Multithreaded Software},
  journal      = {Mathematics},
  volume       = {13},
  number       = {18},
  pages        = {2921},
  year         = {2025},
  doi          = {10.3390/math13182921}
}

@book{wynne2017cucumber,
  author       = {Matt Wynne and Aslak Helles{\o}y and Steve Tooke},
  title        = {The Cucumber Book: Behaviour-Driven Development for Testers and Developers},
  edition      = {2},
  year         = {2017},
  publisher    = {Pragmatic Bookshelf},
  address      = {Raleigh, NC}
}

@article{karlsson2024behaviours,
  author       = {Stefan Karlsson and Robbert Jongeling and Adnan {\v{C}}au{\v{s}}evi{\'c} and Daniel Sundmark},
  title        = {Exploring Behaviours of {RESTful APIs} in an Industrial Setting},
  journal      = {Software Quality Journal},
  volume       = {32},
  number       = {3},
  pages        = {1287--1324},
  year         = {2024},
  doi          = {10.1007/s11219-024-09686-0}
}

@inproceedings{martinlopez2020restest,
  author       = {Alberto Martin-Lopez and Sergio Segura and Antonio Ruiz-Cort{\'e}s},
  title        = {{RESTest}: Black-Box Constraint-Based Testing of {RESTful Web APIs}},
  booktitle    = {Service-Oriented Computing},
  series       = {Lecture Notes in Computer Science},
  volume       = {12571},
  pages        = {459--475},
  year         = {2020},
  publisher    = {Springer International Publishing},
  address      = {Cham, Switzerland},
  doi          = {10.1007/978-3-030-65310-1_33}
}

@article{gorogiannis2019racerd,
  author       = {Nikos Gorogiannis and Peter W. O'Hearn and Ilya Sergey},
  title        = {A True Positives Theorem for a Static Race Detector},
  journal      = {Proceedings of the ACM on Programming Languages},
  volume       = {3},
  number       = {POPL},
  articleno    = {57},
  numpages     = {29},
  year         = {2019},
  doi          = {10.1145/3290370}
}

@inproceedings{vojdani2016goblint,
  author       = {Vesal Vojdani and Kalmer Apinis and Vootele R{\~o}tov and Helmut Seidl and Varmo Vene and Ralf Vogler},
  title        = {Static Race Detection for Device Drivers: The {Goblint} Approach},
  booktitle    = {Proceedings of the 31st IEEE/ACM International Conference on Automated Software Engineering (ASE '16)},
  pages        = {391--402},
  year         = {2016},
  publisher    = {ACM},
  address      = {New York, NY, USA},
  doi          = {10.1145/2970276.2970337}
}

@misc{w3techs2026adobecommerce,
  author       = {{W3Techs}},
  title        = {Usage of {Adobe Commerce} as Content Management System Broken Down by Ranking},
  year         = {2026},
  url          = {https://w3techs.com/technologies/breakdown/cm-magento/ranking},
  note         = {Statistics dated September 5, 2026. Accessed September 5, 2026}
}

@article{beyer2021testcomp,
  author       = {Dirk Beyer},
  title        = {First International Competition on Software Testing},
  journal      = {International Journal on Software Tools for Technology Transfer},
  volume       = {23},
  number       = {6},
  pages        = {833--846},
  year         = {2021},
  publisher    = {Springer},
  doi          = {10.1007/s10009-021-00613-3}
}

@inproceedings{lyu2023miner,
  author    = {Chenyang Lyu and Jiacheng Xu and Shouling Ji and
               Xuhong Zhang and Qinying Wang and Binbin Zhao and
               Gaoning Pan and Wei Cao and Peng Chen and Raheem Beyah},
  title     = {{MINER}: A Hybrid {Data-Driven} Approach for
               {REST} {API} Fuzzing},
  booktitle = {32nd USENIX Security Symposium
               (USENIX Security 23)},
  year      = {2023},
  month     = aug,
  address   = {Anaheim, CA},
  pages     = {4517--4534},
  isbn      = {978-1-939133-37-3},
  publisher = {USENIX Association},
  url       = {https://www.usenix.org/conference/usenixsecurity23/presentation/lyu}
}

@inproceedings{musuvathi2008chess,
  author    = {Madanlal Musuvathi and Shaz Qadeer and Thomas Ball and
               G{\'e}rard Basler and Piramanayagam Arumuga Nainar and
               Iulian Neamtiu},
  title     = {Finding and Reproducing Heisenbugs in Concurrent Programs},
  booktitle = {8th USENIX Symposium on Operating Systems Design and
               Implementation (OSDI 08)},
  year      = {2008},
  address   = {San Diego, CA},
  pages     = {267--280},
  publisher = {USENIX Association},
  url       = {https://www.usenix.org/legacy/event/osdi08/tech/full_papers/musuvathi/musuvathi.pdf}
}

@misc{magentoIssue37465,
  author       = {weissye},
  title        = {Problem when adding an product to the cart in two windows for the same user},
  year         = {2023},
  howpublished = {GitHub issue \#37465, magento/magento2},
  url          = {https://github.com/magento/magento2/issues/37465},
  note         = {Opened May 4, 2023. Accessed: 2026-09-03}
}

@misc{magentoIssue37466,
  author       = {weissye},
  title        = {Shipping fee calculation error in the case of adding an item to the cart during the checkout process},
  year         = {2023},
  howpublished = {GitHub issue \#37466, magento/magento2},
  url          = {https://github.com/magento/magento2/issues/37466},
  note         = {Opened May 4, 2023. Accessed: 2026-09-03}
}

\end{document}